\documentclass[12pt]{iopart}

\usepackage[]{graphicx}

\begin{document}
\title{Reconsidering Rapid Qubit Purification by Feedback}
\author{H.M. Wiseman}
\address{Centre for Quantum Computer Technology, Griffith
University, Brisbane, Queensland 4111 Australia}
\author{J.F. Ralph}
\address{Department of Electrical Engineering and Electronics, The University of Liverpool, 
Brownlow Hill, Liverpool, L69 3GJ, United Kingdom}
\date{\today}

\newcommand{\beq}{\begin{equation}}
\newcommand{\eeq}{\end{equation}}
\newcommand{\beqa}{\begin{eqnarray}}
\newcommand{\eeqan}{\end{eqnarray*}}
\newcommand{\beqan}{\begin{eqnarray*}}
\newcommand{\eeqa}{\end{eqnarray}}
\newcommand{\bra}[1]{\langle{#1}|}
\newcommand{\Bra}[1]{\left\langle{#1}\right|}
\newcommand{\ket}[1]{|{#1}\rangle}
\newcommand{\Ket}[1]{\left|{#1}\right\rangle}
\newcommand{\ip}[1]{\langle{#1}\rangle}
\newcommand{\Ip}[1]{\left\langle{#1}\right\rangle}
\newcommand{\non}{\nonumber}
\newcommand{\id}{\openone}
\newcommand{\eq}[1]{Eq.~(#1)}
\newcommand{\eqr}[1]{Eq.~(\ref{#1})}
\newcommand{\eqrs}[1]{Eqs.~(#1)}
\newtheorem{theo}{Theorem}

\newcommand{\bqa}{\begin{eqnarray}}
\newcommand{\eqa}{\end{eqnarray}}
\newcommand{\nn}{\nonumber}
\newcommand{\nl}[1]{\nn \\ && {#1}\,}
\newcommand{\erf}[1]{Eq.~(\ref{#1})}
\newcommand{\erfs}[2]{Eqs.~(\ref{#1})--(\ref{#2})}
\newcommand{\dg}{^\dagger}
\newcommand{\rt}[1]{\sqrt{#1}\,}
\newcommand{\smallfrac}[2]{\mbox{$\frac{#1}{#2}$}}
\newcommand{\half}{\smallfrac{1}{2}}
\newcommand{\ito}{It\^o }
\newcommand{\str}{Stratonovich }
\newcommand{\du}{\partial}
\newcommand{\dbd}[1]{{\du}/{\du {#1}}}
\newcommand{\sq}[1]{\left[ {#1} \right]}
\newcommand{\cu}[1]{\left\{ {#1} \right\}}
\newcommand{\ro}[1]{\left( {#1} \right)}
\newcommand{\an}[1]{\left\langle{#1}\right\rangle}
\newcommand{\st}[1]{\left|{#1}\right|}
\renewcommand{\tr}[1]{{\rm Tr}\sq{ {#1} }}

\newcommand{\s}[1]{\hat \sigma_{#1}}

\newcommand{\artanh}{{\rm artanh}}

\pacs{03.67.-a, 03.65.Ta, 02.50.Tt, 02.50.Ey}

\begin{abstract}
This paper reconsiders the claimed rapidity of a scheme 
for the purification of the quantum state of a qubit, proposed recently in Jacobs 2003 {\it Phys. Rev. A} {\bf 67} 030301(R). The qubit starts in a completely mixed state, and information is obtained by a continuous measurement. 
Jacobs' rapid purification protocol uses Hamiltonian feedback control to maximise 
the {\em average purity} of the qubit for a {\em given time}, with a factor of two increase in the purification rate over the no-feedback protocol.  However, by re-examining the latter approach, we show that it mininises the {\em average time} taken for a qubit to reach a {\em given purity}. 
 In fact, the average time taken for the no-feedback protocol beats that for Jacobs' protocol by a factor of two. 
 We discuss how this is compatible with Jacobs' result, and the usefulness of the different approaches.
\end{abstract}

\maketitle

\section{Introduction}

Quantum feedback for continuously monitored systems has been shown to be useful for controlling the state of a quantum system \cite{Bel87,Wis93,HorKil97,Doh99,Kor01,Wis02},
and for improved  parameter estimation by adaptive measurement \cite{Dol73,Wis95}. Both applications have been demonstrated in some prominent experiments \cite{Smi02,Arm02,GerStoMab04,Bus06}. Recently, 
Jacobs \cite{Jac03,Jac04}  and co-workers \cite{ComJac06} have proposed an application for quantum feedback control that incorporates aspects of both applications: rapid purification.   As for adaptive measurements, rapid purification uses feedback to increase the amount of information extracted by the measurement. However, this information is about the final state of the quantum system, not  a pre-existing parameter. Also unlike adaptive measurements (but as in state-control by feedback), the feedback in rapid purification does alter the system state on average. However, unlike in state-control,  it is not the goal to produce a particular final state, only a final state of high purity (although further Hamiltonian feedback can always turn the final almost-pure state into any desired almost-pure state). 

In the cases considered \cite{Jac03,Jac04,ComJac06}, the objective of the feedback protocol
was to maximise the rate of increase of the average purity of the quantum state. That is, to maximise the {\em average purity} in a {\em given time}. In the simplest case,
a single qubit \cite{Jac03,Jac04}, this is achieved by continually rotating the Bloch vector onto  
the plane that is orthogonal to the measurement axis. Jacobs showed that this gives rise to a factor 
of two improvement in the rate of purification over the no-feedback case. (Here Jacobs assumes that the measurement basis coincides with the qubit energy basis, so that in the no-feedback case the Bloch vector  remains on the measurement axis.) 

In this paper,
we demonstrate that Jacob's approach, although of undoubted interest and possible value, 
misses some important properties of the purification process. In particular, we show that the opposite approach --- keeping the Bloch vector on the measurement axis --- minimises the {\em average time} for the qubit to reach a {\em given purity}. In fact, we also find a factor of two improvement in the average time over Jacobs' protocol. 
While our results pertain only to qubits, the techniques we employ and the different ways to characterize purification rates in stochastic systems are potentially significant for broad aspects of
quantum information science.

This paper is organised as follows. In Sec.~II we analyse continuous measurement of a qubit with no feedback, for which we use a $c$ subscript because it has a classical analogue in continuous measurement of a bit. We obtain expressions for $\bar{T}_c$, the average time to reach purity $1-\epsilon$, and $\tau_c$, the time for the average purity to reach $1-\epsilon$. (We discuss the operational meaning of the average purity in an Appendix.) 
In Sec.~III we summarize Jacobs' quantum feedback protocol for rapid purification, for which we use a $q$ subscript because it has no classical analogue. We show that, for $\epsilon \ll 1$, $\bar{T}_q = 2\bar{T}_c$ even though $\bar{T}_q = \tau_q = \tau_c/2$. In Sec.~IV we explain how these apparently paradoxical results can occur by considering the (analytically derived) distributions for qubit purities under the different protocols. We also find numerically the full distribution of times taken to reach a given purity. Finally, we give a simple analytical argument for why $\bar{T}_c = \bar{T}_q/2$, which also establishes that the no-feedback protocol is optimal for minimizing $\bar{T}$. We conclude in Sec.~V with a discussion of the implications of our work for initializing quantum registers.   We emphasize that the remarkable result \cite{ComJac06}, that feedback allows an $O(d)$-fold increase in purification rate for a $d$-level system, seems unlikely to be affected by our argument here.

\section{The No-Feedback Protocol}

\subsection{Monitoring a Qubit}

We start by considering the action of continuous measurement 
(which can be thought of as continual weak measurements) on a simple model qubit,
with no Hamiltonian (neither controlled by feedback nor constant).
The conditional evolution equation for the qubit state matrix 
under continuous weak measurement of $\s{z}$ is 
\cite{Jac03,Jac04,Wis96a}
\beq
d\rho = dt {\cal D}[\s{z}]\rho  + dW{\cal H}[\s{z}]\rho.
\eeq
Here, for arbitrary operators $\hat{c}$ and $\rho$, 
\bqa
{\cal D}[\hat c] \rho &\equiv& \hat c\rho \hat c\dg - \frac{1}{2}\{\hat c\dg \hat c,\rho\} \\
{\cal H}[\hat c] \rho &\equiv& \ro{ \hat c\rho - {\rm Tr}[\hat c\rho]\rho } + {\rm H.c.} ,
\eqa
while $dW$ is an infinitesimal Wiener increment satisfying $dW^2=dt$ \cite{Gar85} which 
is the {\em innovation} in the measurement result \cite{Doh99,Wis02}. 
(Note that Jacobs' measurement strength parameter $k$ has the value $1/2$ in our units.)

Defining $z={\rm Tr}[\s{z}\rho]$ {\em etc.} we can represent $\rho$ by the Bloch vector $(x,y,z)$.
By symmetry we can, without loss of generality, take $y=0$. Then the conditional evolution of the qubit is given by 
\bqa \label{SDEz}
dz &=& 2 (1-z^2) dW , \\
dx &=& -2xdt - 2zx dW. \label{SDEx}
\eqa
In terms of these variables, the purity of the qubit is given by 
\beq
p = \tr{\rho^2} = (1+x^2+z^2)/2. \label{purity}
\eeq
If we choose the initial condition $z=z_0$ and $x=0$ at $t=0$, then it is easy to see that $x(t)=0$ for all times. Then the qubit is characterized just by $z(t)$. 

 \subsection{Average Purity}
 
Rather than using \erf{SDEz}, it is convenient to study the probability distribution for $z$ using linear quantum trajectory theory \cite{GoeGra94,Wis96a}. By this method, Jacobs has shown \cite{Jac04} that for a qubit initially in the maximally mixed state ($z_0=0$), 
$z$  is a random variable given by
\beq \label{defq}
z(t) = \tanh(2 q),
\eeq
where $q$ is a random variable having the probability distribution
\beq \label{wpq}
\wp(q)dq = \frac{e^{-2t}}{\sqrt{2\pi t}} \cosh(2q) e^{-q^2/2t}dq.
\eeq 
From this, the average qubit purity at time $t$ is  \cite{Jac03,Jac04}
\beq \label{exactp}
\bar{p}(t) = \int \frac{1+\tanh^2(2 q)}{2}\wp(q)dq  = 1 - \frac{e^{-2t}}{\sqrt{8 \pi t}}\int_{-\infty}^{\infty} \frac{e^{-q^2/2t}}{\cosh(2q)} dq.
 \eeq
  Note that we use $\bar{p}$ for the average purity, not for $1$ minus the purity as Jacobs does \cite{Jac04}. The operational significance of $\bar{p}$ is discussed in the Appendix.
  
Say we are interested in reaching a certain target for the mean purity, $\bar{p}=1-\epsilon$ for $\epsilon \ll 1$. This will occur in the long time limit $t\gg1$, but even for $t = 5$, it is a very good approximation to replace \erf{exactp} with the asymptotic ($t \gg 1$) expression 
 \beq \label{nqreps}
\epsilon \simeq \frac{e^{-2t}}{\sqrt{8 \pi t}}\int_{-\infty}^{\infty} \frac{dq}{\cosh(2q)}  = \frac{\pi e^{-2t}}{4\sqrt{2 \pi t}} 
 \eeq
 Thus, the time $t=\tau_c$ at which the average purity reaches the level $1-\epsilon$ is, 
 in the asymptotic regime of $\epsilon \ll 1$, 
 \beq \label{log}
\tau_c \sim \frac{1}{2}\ln (\epsilon^{-1}) .
 \eeq
Here the subscript $c$ denotes classical, because this measurement involves no 
coherences (it applies equally to a measurement of a classical bit as to a qubit).
This is because, as noted above, 
the action of the measurement of $\s{z}$ does not move the Bloch vector
away from the $z$-axis and we are ignoring any Hamiltonian evolution. 
In more general situations, where the intrinsic 
qubit Hamiltonian does not commute with $\s{z}$, the Bloch vector will move
away from the measurement axis. If this is the case, Hamiltonian feedback control would be
required to rotate the Bloch vector back onto the measurement axis in order to reduce 
the problem to the classical one solved here.

\subsection{Average Purification Time}

Rather than the time taken for the average purity to reach a given level, we may be more interested in the time at which a particular system reaches a target purity $p =  1-\epsilon$. 
For this it is more convenient to return to the (nonlinear) quantum trajectory equation (\ref{SDEz}). 
From this equation, an ensemble of qubits with $z=z_0$ and $x=0$ at $t=0$ is represented by the probability distribution $\wp(z;t|z_0;0)$ obeying the Fokker-Planck equation \cite{Gar85}
\beq
\dot{\wp}(z;t) = \frac{\du^2}{\du z^2} 2(1-z^2)^2 \wp(z;t),
\eeq
with initial condition $\wp(z;0) = \delta(z-z_0)$. We wish to study the time $T$ at which $|z|$ first attains the value $Z$ such that $(1+Z^2)/2 = 1-\epsilon$.  We can do this by considering the above Fokker-Planck equation with absorbing boundary conditions at $Z$ and $-Z$. Writing its solution as $\tilde{\wp}(z,t)$ [where the tilde denotes the presence of boundary conditions], we can thereby define the probability that $T>t$ for an initial value of $z_0$ as \cite{Gar85}
\beq
G(t|z_0)  =  \int_{-Z}^{Z} dz \tilde\wp(z;t|z_0;0).
\eeq

It is simple to see that  the mean time until the qubit is purified to the desired level is given by  
\beq \label{barTfromG}
\bar{T}(z_0) = \int_0^\infty G(t|z_0)dt
\eeq
Following the method in Ref.~\cite{Gar85}, this quantity (known as the average time of first passage) obeys the ordinary differential equation
\beq
2(1-z_0^2)^2 \bar{T}''(z_0) = -1,
\eeq
where the dash represents differentiation with respect to $z_0$. The boundary conditions are $\bar{T}(-Z) = \bar{T}(Z) = 0$, since a qubit with $|z_0|=Z$ is at the boundary already so $\bar{T}=0$ there by definition. The solution to this problem is 
\beq
\bar{T}(z_0) = \frac{1}{4} ( Z\artanh Z - z_0 \artanh z_0).
\eeq
 
 We are again interested in the case where the qubit is initially in a completely mixed state: $z_0 = 0$. Then  $\bar{T} = (1/4)Z\artanh Z$. For $\epsilon$ small we have $Z \simeq 1- \epsilon$, and 
 \beq
 \bar{T}_c \simeq \frac{1}{8} \ln (2/\epsilon) \sim \frac{1}{8} \ln (\epsilon^{-1})
 \eeq
 (Note the subscript $c$ again.) That is, the {\em mean time} $\bar{T}_c$ to reach a purity of $1-\epsilon$ is {\em one quarter} the size of the 
time $\tau_c$ to reach a {\em mean purity} of $1-\epsilon$.

 \section{Jacobs' Protocol}
 
 Jacobs' protocol is to maximize the increase in the average purity at every point in time. From \erfs{SDEx}{purity} it is easy to show (using the \ito calculus \cite{Gar85}) that 
 \beq \label{SDEs}
 ds = -(8s^2 + 4x^2 s)dt - 4zsdW.
 \eeq
 Here $s=1-p$ is sometimes known as the linear entropy. Note the distinction between $s$, a time-dependent random variable, and $\epsilon$, a fixed parameter 
related to a target purity to be achieved by a protocol. It is clear that on average
 $s$ will decrease most rapidly if $x$ is maximized. That is, if one uses feedback to 
 rotate the Bloch vector onto  the $x$ axis. Then one has $z=0$ and $x^2 = 1-2s$, so  
 \beq \label{ODEs}
 ds = -4sdt
 \eeq
 Thus, under Jacobs' rapid purification scheme the evolution under ideal
 Hamiltonian feedback control is deterministic! The purity of the qubit is given by 
 \beq 
 {\rm Tr}[\rho^2] = 1 - \frac{1}{2} e^{-4t} ,
 \eeq
identically for all qubits, so $p = \bar{p}$, and also 
$T= \bar{T} = \tau$. That is, the time of first passage and the time for the average purity to reach the desired level are equal.  For a target purity of $1-\epsilon$, these times are 
 \beq \label{tq}
  \tau_q = T_q = -\frac{1}{4} \ln 2\epsilon \sim \frac{1}{4} \ln(\epsilon^{-1})
 \eeq
 Here the subscript $q$ stands for quantum, since this adaptive technique exploits the qubit coherences and cannot be applied to a classical bit.
 For $\epsilon \ll 1$, $T_q$ is exactly {\em twice} as long as $\bar{T}_c$, 
 even though $\tau_q$ is {\em half} as long as $\tau_c$. 
 
\section{How is this possible?}

\subsection{Distributions of Purity}

Some insight into these results can be found from considering $\wp(p;t)dp$, the probability distribution for the purity at time $t$. In terms of this, the {\em mean} purity at a given time $t$ is $\bar{p}(t) = \int_0^1 p  \wp(p;t)dp$, so the time $\tau$ at which the purity reaches the desired value is defined as
\beq
1-\epsilon = \int_0^1 p  \wp(p;\tau)dp
\eeq
To define $\bar{T}$ it is necessary to consider a different distribution, $\tilde{\wp}(p;\tau)dp$, which arises from the absorbing boundary conditions as described in Sec.II. 
In terms of this, the time $\bar{T}$ is given by 
\beq
\bar{T} = \int_0^\infty dt  \int_{0}^{1-\epsilon} dp \tilde\wp(p;t).
\eeq
(This is derived in the same way as \erf{barTfromG}.) However, when $\epsilon$ is small it is very unlikely (with probability of order $\epsilon$ in fact) for the purity of the qubit to reduce significantly below $1-\epsilon$ once it has crossed that barrier. Thus 
$\wp(p;t)$ is sufficiently similar to $\tilde\wp(p;t)$ 
that it can help us to understand $\bar{T}$ also. 
From \erfs{purity}{wpq}, it follows that for the classical (no-feedback) measurement the distribution of 
purities is 
\beq
\wp_c(p;t)dp = \frac{e^{-2t}}{8\sqrt{\pi t}} 
 \frac{\exp\left(-\smallfrac{1}{8t}\artanh^2\sqrt{2p-1}\right)} {\sqrt{(1-p)^{3}(2p-1)}} dp.
\eeq
By contrast, under Jacobs' quantum feedback protocol 
\beq
\wp_q(p;t)dp = \delta(p - 1 +  e^{-4t}/2)dp
\eeq

\begin{figure}
\begin{center}
\includegraphics[height=12cm]{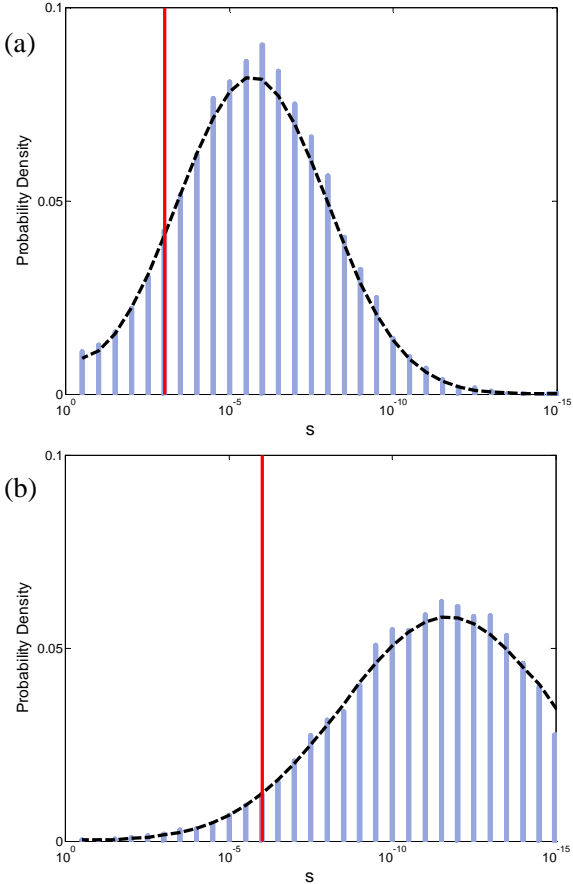}
\end{center}
\caption{Probability distribution of purity values  generated by the no-feedback 
scheme --- theoretical (black dashed line) and numerical (blue histogram) --- 
and Jacobs' deterministic scheme (red solid line) at different times: (a) $t=1.72694$ and (b) $t= 3.2806$. Note the logarithmic scale for $s=1-p$.}
\end{figure}

The two purity distributions are plotted in Fig.~1 for two different times, 
$t=\bar{T}_c$ and $t=T_q=\tau_q$. With $\epsilon = 10^{-6}$, these 
times are $t=\bar{T}_c = 1.72694$ and $t=T_q=\tau_q = 3.2806$. The
agreement between the numerical histograms and theoretical results are good; the 
 slight differences between the two  are related to the limited numbers of runs ({\em circa} 20 000) and rounding errors in the numerical calculations for very low $\epsilon$ values. 
In each case, Fig.~1(a) and 1(b), the majority of the probability distribution for the classical (stochastic) scheme lies to the right of the quantum (deterministic) scheme which is indicated by the red line. This means that the majority of cases should produce a significantly lower value for the purity
than Jacobs' approach, whilst the average purity for the stochastic case is unduly affected 
by the relatively few instances where the purity is still relatively far from one.

\begin{figure}
\begin{center}
\includegraphics[height=6cm]{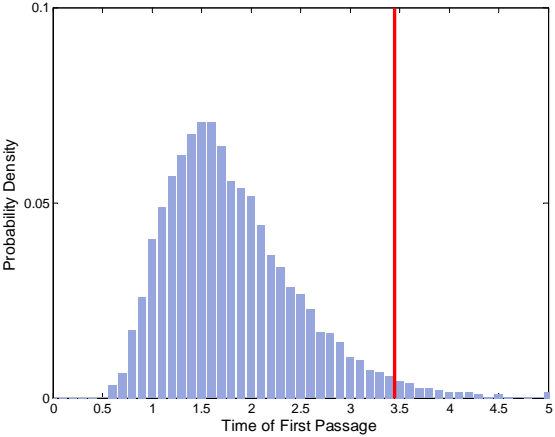}
\end{center}
\caption{Probability distributions of average times of first passage for $\epsilon = 10^{-6}$ for present
scheme (blue histogram) and Jacobs' deterministic scheme (red solid line).}
\end{figure}

\subsection{Distributions of Purification Time}

In Fig.~2 we plot the classical and quantum probability distribution for the time of first passage ($T$), as determined from numerical simulation of (\ref{SDEz}) and (for Jacobs' protocol) from \erf{tq}, for the same value of $\epsilon$. As we would expect from Fig.~1, the majority of classical (stochastic) cases reach the desired purity 
level before they would using the deterministic, ideal feedback scheme.
In Fig.~3(a) we plot the average first passage time as a function of $\epsilon$, and in Fig.~3(b) the ratio of average first passage times produced by two measurement schemes as a function of $\epsilon$. Fig.~3(b) shows that the relative benefit of the classical approach tends to the theoretical value of $T_q/\bar{T}_c \rightarrow 2$ 
as $\epsilon \rightarrow 0$. Note that the  
improvement offered by our classical protocol is significant even for quite low target purities (e.g. $\epsilon=0.01$). Moreover,  Jacobs' purification time and the average first passage time are separated by at least one standard deviation of the first passage time for all $\epsilon \leq 0.003$. These facts are encouraging for experimentally demonstrating the differences between the schemes, even
if Jacobs' ideal Hamiltonian feedback scheme cannot be realized perfectly in the laboratory.
\begin{figure}
\begin{center}
\includegraphics[height=11cm]{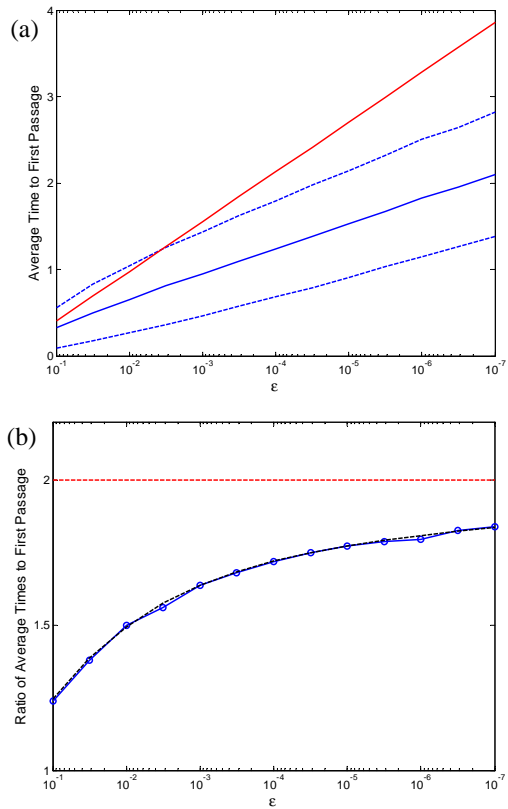}
\end{center}
\caption{(a) Average time to first passage for Jacobs' scheme (solid red line) and 
the scheme presented here (solid blue line) with dashed lines indicating standard deviation of the
first passage distribution for $T_c$,   (b) Ratio of average first passage times $T_q/\bar{T}_c$:  data joined by solid line; theoretical curve $\ln(1/2\epsilon)/\sqrt{1-2\epsilon}\,\artanh \sqrt{1-2\epsilon}$ (black dashed line which asymptotes to 2 and is partially covered by numerical results). Both graphs are plotted as functions of $\epsilon$ on a log scale. }
\end{figure}

In Jacobs' scheme, the purity of the state always increases. The observer has no prior knowledge about the observable $\s{z}$ being measured because the state is always rotated to lie on the horizontal plane of the sphere. As a consequence, any result always increases the observer's certainty about the state.

By contrast, in the classical scheme analysed here, the measurement may increase or decrease the observer's certainty about the system. Consider the situation where the qubit is in a mixture of being in state $\s{z}=1$ and $\s{z}=-1$, but more likely to be in the former so that $z>0$. 
From \erf{SDEz}, if $dW>0$ in the next infinitesimal interval, then the observer will become more certain that $\s{z}=1$ is the true state of the qubit, and $z$ (the observer's expectation value for $\s{z}$) will increase. On the other hand, if $dW<0$, the observer will become less certain and $z$ will decrease. That is, in any infinitesimal interval, the purity decreases half the time! Why then does the purity increase at all? The answer is that $p$ varies as $z^2$, so even though the positive and negative changes in $z$ balance, the change in $p$ is positive on average. But, not surprisingly, the increase in $\bar{p}$ is slower than under Jacobs' scheme. Also, in the classical case there is clearly going to be a far greater spread in the purity than in Jacobs' scheme, as shown in Fig~1. But this is exactly what allows for the average time $\bar{T}$  to reach a purity threshold to be shorter in the former case.

\subsection{Simple Derivation and Optimality}

The results already presented in this section confirm that the no-feedback protocol gives an improvement over Jacobs' protocol for the average time to purify a qubit, and give some insight into how this is possible. However, what is lacking is (a) a simple derivation of the factor of two improvement, and (b) a demonstration that the no-feedback protocol is optimal for this purpose. We now provide both of these. 

By considering the average of \erf{SDEs} it was easy to see that Jacobs' protocol maximized the decrease in the average linear entropy $\bar{s} = 1-\bar{p}$. However, as argued above, 
for the no-feedback case this average gives undue weight to the tail of the distribution in which $s$ is much larger than the mean. This can be rectified by considering $\ln s$, the distribution of which is much more symmetrical, as seen in Fig.~1. From \erf{SDEs}, the stochastic differential equation for $\ln s$ is, using the \ito calculus \cite{Gar85}, 
\beq
d\ln s = -4 (2s + x^2 + 2z^2)dt  - 4zdW
\eeq
Taking the average of this equation, we see that the  most rapid purification (for a given $s$) is  when $z^2$ is maximized at $1-2s$ and $x^2$ set to zero; that is, exactly the classical protocol considered above. Moreover, in this classical case we obtain, for almost-pure states ($z^2 \simeq 1$), 
\beq \label{lns1}
d \an{\ln s}/dt \simeq - 8,
\eeq
which is the characteristic rate for purification to a high level of purity. 
By contrast, under Jacob's protocol we set $z^2=0$ and $x^2=1-2s$ to obtain 
\beq
d \ln s/dt = - 4.
\eeq
This is of course the rate of purification found in \erf{ODEs}, and is slower than the classical  rate (\ref{lns1}) by the factor of two found previously.

\section{Conclusions}

In this paper, we have considered some interesting properties of the rapid purification protocol
proposed by Jacobs for quantum systems subject to continuous  measurements and Hamiltonian
feedback control. In particular, we have demonstrated that, although Jacobs' scheme (keeping the Bloch vector perpendicular to the measurement axis) provides the fastest increase in the average purity of the qubit,  the opposite approach (keeping the Bloch vector on the measurement axis) provides the shortest average time to reach a given purity level.  This counter-intuitive result is due to the following: in Jacobs' approach the purification
is deterministic, whilst the classical approach is stochastic where the distribution of purities at a given time is heavily skewed. Thus, although 
 the majority of qubits will reach a given purity level
quicker than under Jacobs' scheme, there is a minority of qubits that have a relatively low purity,
 which reduces the expected (average) purity below that of Jacobs' scheme. 

Aside from their theoretical interest, these results have most obvious application in the initialization of qubits for quantum information processing. The fact that, under the classical protocol, half the qubits  end up in the $\s{z}=1$ state and half in the $\s{z}=-1$ state does not matter; it is known which state the qubit is in so it can always be rotated to the desired fiducial state. Since it is increasingly common to consider non-deterministic protocols in quantum computing, it seems likely that a scheme that minimizes the average time to purify the qubits in an ensemble would be preferred. The fact that no  Hamiltonian feedback control is required for the classical protocol is another advantage it has over Jacob's protocol. However, if it were necessary to purify {\em all} members of an ensemble of qubits to a given degree of purity in as short a time as possible, then Jacobs' rapid purification by quantum feedback should be used.

Finally, we wish to emphasize the following. In very recent work, Combes and Jacobs \cite{ComJac06} have shown that Hamiltonian feedback control can be used to purify a qudit (that is, a $d$-level quantum system) a factor of order $d$ faster than can be achieved without feedback. There is nothing in our analysis to suggest that this factor of $d$ would change if one calculated the average time to purify (as we considered here) rather than the average purity, as Jacobs has always considered \cite{Jac03,Jac04,ComJac06}. We suggest that any difference between the two types of purification time would be of order unity, but that remains to be confirmed by future work.

\section*{ACKNOWLEDGMENTS}
JFR would like to thank the UK Engineering and Physical
Science Research Council (Grant No.: EP/C012674/1) for their generous support of
this work. HMW thanks the Australian Research Council and the Queensland Government.
We both acknowledge friendly discussions with Kurt Jacobs. 

\section*{Appendix}

One could question whether the average purity $\bar{p}$ (\ref{exactp}) as used by Jacobs 
is a meaningful concept operationally. If one were to monitor a qubit (or bit) 
as above, then forget the particular measurement record obtained, and then ask what 
is the purity of the resulting state, the answer is not given by \erf{exactp}.  
The reason is that purity is a nonlinear function of the state, unlike $z$ for example. 
Of course if one were to ignore completely the measurement record then the purity of the resultant state would be  $1/2$, because the state is as likely to be purified towards $z=1$ as to $z=-1$. 
Thus it is only sensible to say that at the end of the measurement the state is first 
rotated (if necessary) so that $z>0$, and then the measurement record is forgotten.
The resultant state is then defined by the expected value of $\s{z}$, which in terms of the 
unrotated stochastic variable $z$ is given by $\an{|z|}$. From above, this evaluates to
\beq
\an{|z|} = 2\int_0^{\infty} \sinh (2q) \wp(q) dq = 
1 - \frac{2e^{-2t}}{\sqrt{2 \pi t}}\int_{0}^{\infty} {e^{-q^2/2t-2q}} dq.
\eeq
The purity of this average state is $p_{\rm av.} = 1 + \an{|z|}^2$. 
Putting $p_{\rm av} = 1-\epsilon$, in the limit $t\gg 1$ this evaluates to 
\beq
\epsilon \simeq \frac{2e^{-2t}}{\sqrt{2 \pi t}}\int_{0}^{\infty} {e^{-2q}} dq = 
\frac{e^{-2t}}{\sqrt{2 \pi t}} .
 \eeq
This is only a constant factor ($4/\pi$) larger than that calculated above (\ref{nqreps}),
so the asymptotic expression for $\tau_c$ in \erf{log} is unchanged. For these reasons, 
in the body of this paper we just use Jacobs' $\bar{p}$ for simplicity.


\section*{References}
\begin{thebibliography}{99}

\bibitem{Bel87}
Belavkin V P 1987, ``Non-demolition measurement and control in quantum dynamical systems'', in {\em Information, complexity, and control in quantum physics}, edited by A. Blaqui\`ere, S. Dinar, G. Lochak (Springer, New-York); Belavkin V P 1999 {\em Rep. Math. Phys.} {\bf 45} 353, and references contained therein.
\bibitem{Wis93} 
Wiseman H M and Milburn G J 1993 {\it Phys. Rev. Lett.} {\bf 70} 548; Wiseman H M 1994 {\em Phys. Rev. A} {\bf 49} 2133.
\bibitem{HorKil97}
Mabuchi H and  Zoller P 1996 
{\it Phys. Rev. Lett.} {\bf 76} 3108; 
Horoshko D B and Kilin S Ya 1997 {\it Phys. Rev. Lett.} {\bf 78} 840.
\bibitem{Doh99}
Doherty A C and Jacobs K 1999 {\it Phys. Rev. A} {\bf 60} 2700; 
Doherty A C, Habib S, Jacobs K, Mabuchi H and Tan S M 2000 {\it Phys. Rev. A} {\bf 62} 012105.
\bibitem{Kor01}
Korotkov A N 2001 {\it Phys. Rev. B} {\bf 63} 115403; Korotkov A N 2005 {\it Phys. Rev. B} {\bf 71} 201305(R). 
\bibitem{Wis02}
Wiseman H M, Mancini S and Wang J 2002 {\it Phys. Rev. A} {\bf 66} 013807; Wiseman H M and Doherty A C 2005 {\it Phys. Rev. Lett.} {\bf 94} 070405.
\bibitem{Dol73}
Dolinar S J 1973 Research Laboratory of Electronics, MIT, Quarterly Progress Report {\bf 111}, p.~115. 
See also: Helstrom C W 1976 {\em Quantum Detection and Estimation Theory}, p.~163 (Academic, New York).
\bibitem{Wis95}
Wiseman H M 1995 {\it Phys. Rev. Lett.} {\bf 75} 4587; 
Berry D W and Wiseman H M 2000 {\it Phys. Rev. Lett.} {\bf 85} 5098.
\bibitem{Smi02}
Smith W P, Reiner J E, Orozco L A, Kuhr S and Wiseman H M 2002 {\it Phys. Rev. Lett.} {\bf 89} 133601.
\bibitem{Arm02} Armen M A, Au J K, Stockton J K, Doherty A C and Mabuchi H 2002
{\it Phys. Rev. Lett.} {\bf 89}, 133602.
\bibitem{GerStoMab04}  Geremia J M, Stockton J K and Mabuchi H 2004 {\it Science} {\bf 304} 270.
\bibitem{Bus06}
Bushev P {\em et al.} 2006 
{\it Phys. Rev. Lett.} {\bf 96}, 043003  
\bibitem{Jac03}
Jacobs K 2003 {\em Phys. Rev. A} {\bf 67} 030301(R).
\bibitem{Jac04}
Jacobs K 2004 {\it Proc. of SPIE} {\bf 5468} 355.
\bibitem{ComJac06}
Combes J and Jacobs K 2006 {\it Phys. Rev. Lett.} {\bf 96} 010504.
\bibitem{Wis96a}
Wiseman H M 1996 {\it Quantum Semiclass. Opt.} {\bf 8}, 205.
\bibitem{Gar85}
Gardiner C W 1985 {\em Handbook of Stochastic Methods} (2nd edition)(Spring\-er, Berlin).
\bibitem{GoeGra94}
Goetsch P and Graham R 1994 {\it Phys. Rev. A} {\bf 50}, 5242.











\end{thebibliography}
\end{document}